\documentclass[twocolumn,preprintnumbers,showpacs,aps,pra,superscriptaddress]{revtex4}

\AtBeginDocument{
\heavyrulewidth=.08em
\lightrulewidth=.05em
\cmidrulewidth=.03em
\belowrulesep=.65ex
\belowbottomsep=0pt
\aboverulesep=.4ex
\abovetopsep=0pt
\cmidrulesep=\doublerulesep
\cmidrulekern=.5em
\defaultaddspace=.5em
}

\usepackage{amsmath,amssymb}
\usepackage{graphicx}
\usepackage[usenames,dvipsnames]{xcolor}
\usepackage{amsmath, amsthm}
\usepackage{booktabs}
\usepackage{hyperref}
\hypersetup{colorlinks=true,citecolor=Orange,linkcolor=Red,urlcolor=Black}
\begin{document}

\title{Unconventional fermionic pairing states in a monochromatically tilted optical lattice}

\author{A. Nocera}
\affiliation{Computer Science and Mathematics %
Division and Center for Nanophase Materials Sciences, Oak Ridge National Laboratory, %
 \mbox{Oak Ridge, Tennessee 37831}, USA}

\author{A. Polkovnikov}

\affiliation{Department of Physics, Boston University, Boston, Massachusetts 02215, USA}

\author{A.E. Feiguin}

\affiliation{Department of Physics, Northeastern University, Boston, Massachusetts 02115, USA }

\begin{abstract}
We study the one-dimensional attractive Fermionic Hubbard model under the influence of 
 periodic driving with the time-dependent density matrix renormalization
group method. We show that the system can be driven into an unconventional pairing state
characterized by a condensate made of Cooper-pairs with a finite center-of-mass momentum
similar to a Fulde-Ferrell state. We obtain results both in the laboratory and the
rotating reference frames demonstrating that the momentum of the condensate can be
finely tuned by changing the ratio between the amplitude and the frequency of the
driving. In particular, by quenching this ratio to the value corresponding to suppression of
the tunnelling and the Coulomb interaction strength to zero, we are able to
``freeze'' the condensate. We finally study the effects of different initial conditions,
and compare our numerical results to those obtained from a time-independent Floquet
theory in the large frequency regime. Our work offers the possibility of engineering 
and controlling unconventional pairing states in fermionic condensates. 
\end{abstract}

\maketitle
\newpage

\section{Introduction}
Cold atoms in optical lattices\cite{re:Bloch2008,re:Lewenstein2012} currently
represent one of the best candidates for the quantum simulation
\cite{re:Buluta2009,re:Bloch2012,re:Georgescu2014} of bosonic and fermionic systems, given the
extreme level of tunability of the microscopic Hamiltonian
parameters, such as tunneling and interactions. Recent
experimental advances allow for site-resolved detection
 measurements\cite{re:Bakr2010,re:Sherson2010}, the simulation
of artificial gauge fields\cite{re:jaksch2003creation,re:lin2009synthetic,re:Dalibard2011,re:Struck2012}, 
and also the possibility to control  
the behavior of the optical lattice under the influence of high
frequency electromagnetic fields\cite{re:Lignier2007,re:Zenesini2009}. 
In addition, intriguing phenomena such as 
Bloch oscillations and
the formation of repulsively bound pairs have been observed\cite{re:Winkler2006}. 
A particularly interesting phenomenon is provided by the coherent
 destruction of tunneling, in which
driving the system with an oscillating field has the effect of
renormalizing the intersite tunneling amplitude. By tuning the amplitude
 of the driving field, it has been theoretically and experimentally 
demonstrated that one can induce the superfluid-Mott insulator transition\cite{re:eckardt2005superfluid}.

Theoretically, it has been shown that new exotic interactions and
correlated hopping terms\cite{re:Diliberto2011} can 
be effectively generated by the external electromagnetic fields.
These achievements have opened a new line of research, termed Floquet
engineering. Indeed, both
bosonic and fermionic Hubbard models are currently
theoretically and experimentally investigated, with the possibility of 
exploring new exciting physics out of equilibrium.

In Refs.~[\onlinecite{re:Creffield2008,re:Creffield2011}], it has been demonstrated that a high-frequency monochomatic external field can be used not only 
to control the hopping amplitude of cold atoms trapped in a lattice, 
but also the phase of their wave function. In a one dimensional optical lattice of bosonic 
atoms described by a Bosonic Hubbard model, the authors showed that 
a coherent matter current can be generated. The current is induced 
by a shift in the momentum distribution function's central peak, which is proportional to the amplitude of the driving force.

In this paper, we apply the same periodic perturbation to a 
fermionic optical lattice described by the Fermi-Hubbard model. 
We explore the particular regime of parameters where there 
is an attractive interaction between the fermionic particles and 
explore the possibility of realizing and 
manipulating exotic pairing phases.
In this context, Fulde-Ferrell (FF) and Larkin-Ovchinnikov (LO) phases
recently attracted a great deal of interest from both the experimental\cite{re:Zwierlein2006fermionic,re:Partridge2006pairing,re:Liao2010spin}
and the theoretical community\cite{re:Orso2007,re:Feiguin2007,re:Casula2008,re:Luscher2008,re:Rizzi2008,re:Tezuka2008,re:Feiguin2012bcs,re:Guan2013}. 
In the original idea of Fulde and Ferrell, Cooper pairs form 
with finite center-of-mass momentum\cite{re:FuldeFerrel1964}. Larkin and Ovchinnikov 
formulated a related proposal in which the superconducting order parameter 
oscillates in space\cite{re:Larkin1964}. 
These ideas are closely related, because the inohomogeneus order parameter\cite{re:Yang2001,re:Casalbuoni2004} 
may be interpreted as an interference pattern between condensates 
with opposite center-of-mass momenta. In our case, we find a pairing state similar to a FF state, 
consisting of a condensate which moves as a coherent matter field 
of Cooper-like pairs. In particular, we find that the center-of-mass momentum 
of the condensate can be tuned by changing the amplitude and the frequency 
of the external drive.

Using the Floquet 
formalism\cite{re:Bukov2014,re:Bukov2014a,re:Eckardt2015,re:Bukov2015rev,re:Mikami2016}, 
we also derive an effective time-independent Hamiltonian for the system in the presence 
of a high-frequency driving force, and compare the results coming from 
this effective theory by solving the model numerically using the time 
dependent Density Matrix Renormalization Group method (tDMRG) \cite{tdmrg1,tdmrg2}.

We finally investigate the pairing properties of the system 
by using different quenching protocols. We propose to abruptly change 
either the interaction between the fermions, 
or the parameters of the external driving potential independently.
These protocols are investigated numerically using tDMRG and 
could be experimentally accessible in current experimental setups.

The paper is organized as follows. In Section II we present the model 
in different reference frames and apply Floquet theory in the rotating frame. 
In Section III we show that the numerical results 
obtained with tDMRG are in agreement with Floquet theory.
In Section IV we explore different quenching protocols for 
the characterization of the exotic pairing phase found, and we finally summarize our results in the conclusions.

\section{MODEL}

In this work, we consider a one dimensional Fermionic Hubbard chain 
subjected to an external time dependent perturbation. 
The Hamiltonian in the laboratory reference frame is given by 
$H_{lab}(t)=H+V(t)$, where $H$ is given by the 
one dimensional Fermi-Hubbard model 
\begin{equation}\label{HH}
H=-J \sum\limits_{\langle i,j\rangle, \sigma}(c^\dagger_{i,\sigma}
c_{j,\sigma} + h.c.) + U \sum\limits_{i} n_{i,\uparrow}
n_{i,\downarrow},
\end{equation}
while the tilting perturbation is given by

\begin{equation}
\label{Shak}
V(t) = \theta(t) A \sin(\omega t+\theta) \sum\limits_{i,\sigma} 
i n_{i,\sigma},
\end{equation}
and constitutes a linear potential ramp applied to the sites of 
the lattice suddenly turned on at time $t=0$. Notice that this 
driving protocol is different from the one adopted in 
Ref.~[\onlinecite{re:Creffield2008}], where the amplitude was linearly ramped on in time.
In this paper, we assume that the tilting perturbation 
is sinusoidal, with phase $\theta=0$, amplitude $A$ and frequency $\omega$.
In this case, as was observed in Ref.~[\onlinecite{re:Creffield2008}] and as will be clear below, 
the effects of the phase of the drive are maximum, giving a non 
trivial contribution in the effective Floquet description. 
In the Eq.~(\ref{HH}), $J$ is the hopping amplitude between 
nearest neighbor sites
(indicated by $\langle i,j\rangle$), $U$ parametrizes the repulsion between atoms, 
and $c^\dagger_{i,\sigma}$ ($c_{i,\sigma}$) is the standard
fermionic creation (annihilation) operator on site $i$ with spin
${\sigma}$ ($\bar{\sigma}$ indicates the opposite of $\sigma$),
while $n_{i,\sigma}=c^\dagger_{i,\sigma} c_{i,\sigma}$ is the fermionic
occupation operator. The Planck constant is set to
$\hbar=1$, the lattice parameter $a=1$, and all of the energies
are in the units of the hopping $J$. These definitions 
automatically set the time unit to $\hbar/J$.
Throughout this paper we consider an attractive interaction between 
the fermions $U<0$, a lattice with open boundary conditions and filling $N/L=2/3$, where $N$ is the
 total number of fermions and $L$ is the system size.
It is well known that the low energy physics of the model (\ref{HH}) 
can be described in terms of a 
gapped spin mode and a gapless charge mode. The system belongs to the Luther-Emery
universality class and superconducting pair correlations are dominant, 
\begin{equation}\label{Pair}
P_{i,j}=\langle \Delta_i^{\dag} \Delta_j\rangle \propto |i-j|^{-1/K_{\rho}}
\end{equation}
with an non-universal power law decay with exponent $K_{\rho}$ 
determined by the interaction $U$. The main purpose 
of this paper is to investigate the properties of the pairing 
correlation function in Eq.~(\ref{Pair}) in the presence of the 
driving (\ref{Shak}). 

\subsection{Laboratory, rotating reference frames, and Floquet approximation}
It is useful to remind how an observable quantity is calculated in 
the laboratory reference frame. At time $t=0$, the ground state 
$|\Psi_{0}\rangle$ of $H$ is initially calculated. Afterward, 
the time evolved state $|\Psi_{lab}(t)\rangle= U_{lab}(t,0)
|\Psi_{0}\rangle$, where $U_{lab}(t,0)=\mathcal{T}\Big[\textrm{exp}(-i\int_{0}^{t}H_{lab}(t')dt')\Big]$,  is computed under the action of the entire 
time dependent Hamiltonian $H_{lab}(t)$, $\mathcal{T}$ being the time-ordering operator. Assuming that the ground state
is normalized, an observable $O$ is 
then calculated computing the expectation value
\begin{equation}\label{Obslab}
\langle O \rangle_{lab} = \langle \Psi_{lab}(t) |O| \Psi_{lab}(t)\rangle.
\end{equation}

We now consider the formulation of the problem in the ``rotating''
reference frame. We apply to the Hamitonian $H_{lab}(t)$ the
following time dependent canonical transformation 
\begin{equation}
R(t)=e^{-i f(t) \sum_{j}j {\hat n}_{j}}, 
\end{equation}
where 
\[
f(t)=A\int_{0}^{t}d\tau\, \text{sin}(\omega\tau)=A{1-\cos(\omega t)\over \omega}.
\]
In this way, the rotated Hamiltonian $H_{rot}(t)=R^{\dag}(t)H_{lab}(t)R(t)-iR^{\dag}(t)\partial_{t}R(t)$ is still time dependent and assumes the form 
\begin{align}\label{Hrot}
H_{rot}(t)&=-J\sum\limits_{\langle i,j\rangle, \sigma}(e^{-i f(t)} c^\dagger_{i,\sigma}
c_{j,\sigma} + h.c.) \nonumber\\
&+ U \sum\limits_{i} n_{i,\uparrow}
n_{i,\downarrow},
\end{align}
where the hopping term acquires a complex nontrivial phase, which is the 
central quantity in this paper. It is clear
 that, at $t=0$, $H_{lab}(0)=H_{rot}(0)$. Therefore, once the ground  state 
 $|\Psi_{0}\rangle$ is computed, the time evolved state in the rotating frame, 
 $|\Psi_{rot}(t)\rangle$, which can be obtained at each time $t$ from the laboratory 
frame state vector 
 with $|\Psi_{rot}(t)\rangle=R(t)|\Psi_{lab}(t)\rangle$ reads $|\Psi_{rot}(t)\rangle= 
 U_{rot}(t,0)
|\Psi_{0}\rangle$, where $U_{rot}(t,0)=\mathcal{T}\Big[\textrm{exp}(-i\int_{0}^{t}H_{rot}(t')dt')\Big]$. 
An observable can be obtained by evaluating the following expression
\begin{equation}\label{Obsrot}
\langle O \rangle_{rot} = \langle \Psi_{rot}(t) |R(t)OR^{\dag}(t)| \Psi_{rot}(t)\rangle,
\end{equation}
where the observable operator should be correctly 
transformed to the rotated frame.
We now apply Floquet theory to the rotated Hamiltonian Eq.~(\ref{Hrot}). 
According to Floquet theorem, the time evolution operator
$U(t_1,t_2)=\mathcal{T}\Big[\textrm{exp}(-i\int_{t_1}^{t_2}H(t)dt)\Big]$
of any time-periodic Hamiltonian $H(t)=H(t+T)$ can be written 
in the form
\begin{equation}
U(t_1,t_2)=\textrm{exp}[-iK_{F}(t_2)]\textrm{exp}[-iH_{F}(t_2-t_1)]
\textrm{exp}[iK_{F}(t_1)],
\end{equation}
where $H_F$ is an effective time independent Hamiltonian
describing the long time dynamics of the system, called the Floquet 
Hamiltonian, while $K_{F}(t)=K_{F}(t+T)$ is a time-periodic operator (called 
kick operator) describing the fast evolution in one period of time $T=2\pi/\omega$.
It is well known that in the limit of very large frequency or small period 
one can calculate the Floquet Hamiltonian and the kick operator perturbatively using a high frequency expansion. In this paper, we use a particular van Vleck expansion, which is invariant under the choice of the driving phase {\em in the rotating frame} ~\cite{re:Bukov2014,re:Eckardt2015,re:Mikami2016}:
\begin{eqnarray}
H_{F}^{(0)}&=&H_{0}=\frac{1}{T}\int_{0}^{T}H(t)dt\\
H_{F}^{(1)}&=&\frac{1}{\omega}\sum_{m=1}^{\infty}\frac{1}{m}[H_{m},H_{-m}]=\\
&=&\frac{1}{2!T i}\int_{0}^{T}dt_{1}\int_{0}^{t_1}dt_{2}f(t_1-t_2)[H(t_1),H(t_2)]\nonumber \\
K_{F}^{(0)}(t)&=&0\\
K_{F}^{(1)}(t)&=&\frac{1}{i\omega}\sum_{m\neq0} \frac{e^{i m \omega t}}{m} H_{m}
\nonumber \\ &=&-\frac{1}{2}\int_{t}^{t+T}dt'g(t-t')H(t'),
\end{eqnarray}
where we decomposed the Hamiltonian in a Fourier series
 $H(t)=\sum_{m=-\infty}^{\infty}H_{m}e^{i m \omega t}$, with $H_{m}$ 
operator-valued coefficients; the functions $f$ and $g$ are defined as $f(x)\equiv(1-2x/T)$ 
and $g(x)\equiv(1+2x/T)$.
Recently, the importance of the kick operators in the 
stroboscopic and non-stroboscopic dynamics of Floquet systems has been highlighted 
by several authors\cite{re:Bukov2014,re:Bukov2014a,re:Eckardt2015,re:Bukov2015rev}. 
For the rotated Hamiltonian in Eq.~(\ref{Hrot}), the zero-order Floquet Hamiltonian
is given by
\begin{equation}\label{HF}
\begin{aligned}
H_{F}^{(0)}&=-J J_{0}(A/\omega)\sum\limits_{\langle i,j\rangle, 
\sigma}\Big(e^{-i\frac{A}{\omega}}c^\dagger_{i,\sigma}
c_{j,\sigma} + h.c.\Big) \\
&+U \sum\limits_{i} n_{i,\uparrow}
n_{i,\downarrow},
\end{aligned}
\end{equation}
where the amplitude of the hopping matrix element is renormalized by the Bessel function 
$J_{0}(A/\omega)$. Notice that the hopping in the Floquet 
Hamiltonian is complex acquiring a phase $\phi=-A/\omega$. 
This phase comes from transformation to the rotating frame and would be 
absent for the cosinusoidal driving field~\cite{re:Creffield2008}, 
which is equivalent to the shift in the rotating frame. The above equation 
has been derived using the Jacobi-Anger expansion
\begin{equation}
\textrm{exp}\Bigg[i \frac{A}{\omega} \cos(\omega t)\Bigg] = \sum_{m=-\infty}^{\infty} 
i^{m}J_{m}(A/\omega) e^{i m \omega t},
\end{equation}
which allows one to simply calculate the $m$-th component of the Fourier series 
operator-valued coefficients of the Hamiltonian
\begin{align}\label{eq:FourierDec}
H_{m}&=-J J_{m}(A/\omega) \sum\limits_{\langle i,j\rangle, 
\sigma} \Big(e^{-i\frac{A}{\omega}}i^{m} c^\dagger_{i,\sigma}c_{j,\sigma}
+ h.c.\Big)\nonumber\\&+\delta_{m,0}U \sum\limits_{i} n_{i,\uparrow}n_{i,\downarrow},
\end{align}
where it is important to point out that the interaction part contributes only for $m=0$.
Since at zero order of the high frequency expansion the kick operator is zero,
the evolution operator in the rotating frame $U_{rot}(t,0)$ becomes simply
\begin{equation}\label{eq:Urot0}
U_{rot}^{(0)}(t,0)=\textrm{exp}(-i H_{F}^{(0)}t).
\end{equation}
Using the Fourier decomposition in Eq.~(\ref{eq:FourierDec}), it is easy to prove that
at first order one has
\begin{align}
H_{F}^{(1)}&=0,\nonumber\\
K_{F}^{(1)}(t)&=\Big(i {J\over \omega} e^{-i A/\omega}\sum_{m\neq 0} e^{i m \omega t} \times\nonumber\\
&i^m\frac{J_{m}(A/\omega)}{m}\sum\limits_{\langle i,j\rangle, 
\sigma}c^\dagger_{i,\sigma}c_{j,\sigma}\Big) +h.c. \label{eq:kick}
\end{align}
We see that the leading correction to the kick operator is proportional to $J/\omega$.
The time evolution operator $U_{rot}(t,0)$ at the first order of the high 
frequency expansion reads
\begin{equation}\label{eq:Urot1}
U_{rot}^{(1)}(t,0)=\textrm{exp}[-iK^{(1)}_{F}(t)]\textrm{exp}(-i H_{F}^{(0)}t)
\textrm{exp}[iK^{(1)}_{F}(0)],
\end{equation}
with a nontrivial contribution coming from the kick operator. In particular, this contribution is nonzero at any time $t$ implying that there is always a non-zero kick to the initial wave function. Using the Floquet approximation, an observable can be obtained by evaluating the following expression
\begin{equation}\label{ObsFlo}
\langle O \rangle_{F} = \langle \Psi_{F}(t) |R(t)OR^{\dag}(t)| \Psi_{F}(t)\rangle
\end{equation}
where $|\Psi_{F}(t)\rangle=U_{rot}^{(k)}(t,0)|\Psi_{0}\rangle$, and $U_{rot}^{(k)}(t,0)$
should be calculated at the $k$th order of the high frequency expansion.
Notice that, as for Eq.~(\ref{Obsrot}) the observable operator should be correctly 
transformed to the rotated frame.

\section{Numerical results}

\begin{figure}[h]
\centering \includegraphics[width=8cm]{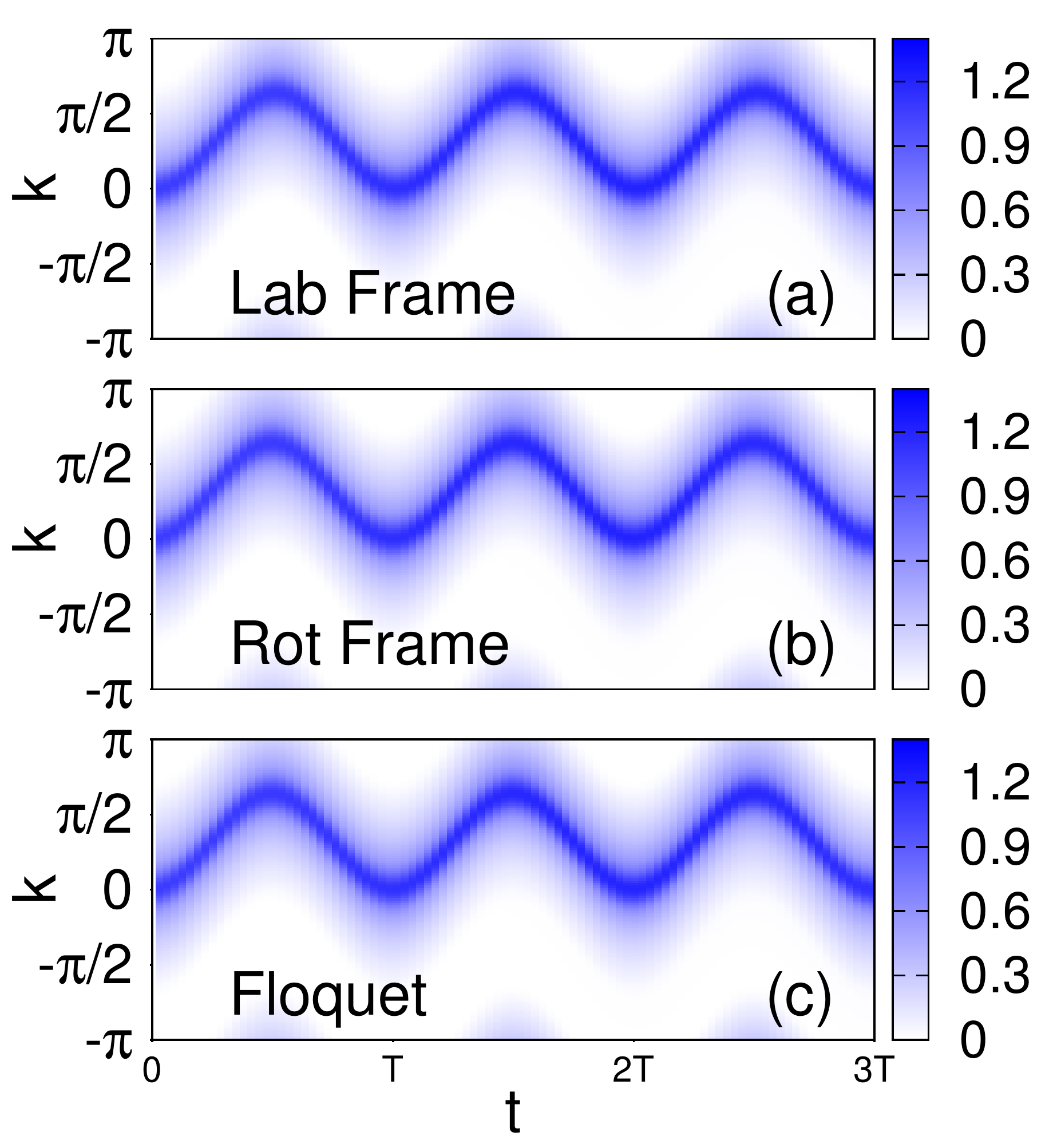}
\caption{(Color online) Pair correlations functions as a function of 
momentum and time in the laboratory, reference frames, and zero-order Floquet 
approximation. In this plot, a $L=12$ sites chain has been simulated using $m=400$ 
DMRG states, $U=-4$ and $A/\omega=0.5$.} \label{fig1}
\end{figure}

We start considering an attractive interaction with strength $U=-4$, a large driving 
frequency $\omega=20$, and an amplitude $A/\omega=0.5$.
Fig.\ref{fig1} shows the Fourier transform of the pair correlation function
defined in Eq.~(\ref{Pair}) as a function of time 
\begin{equation}
P(k,t)=\frac{1}{L}\sum_{i,j}e^{ik(i-j)} \langle\Delta_i^{\dag} \Delta_j\rangle(t).
\end{equation}
Panel (a) is calculated in the laboratory reference frame with  
Eq.~(\ref{Obslab}); in panel (b), Eq.~(\ref{Obsrot}) is used, while 
in panel (c) we use the zero-order Floquet approximation as described in the 
previous section, see Eq.~(\ref{ObsFlo}) where Eq.~(\ref{eq:Urot0}) has been used. 
Notice that, at first order of the Floquet expansion, the only 
extra contribution to the evolution operator comes from the kick operator 
$K_{F}^{(1)}(t)$.  
In the numerical simulations, the ground state $|\Psi_0\rangle$ of $H_{lab}(t)$ at $t=0$ 
is calculated using ground state DMRG. 
We then use a third order Suzuki-Trotter decomposition of 
the time evolution operator,\cite{re:Feiguin2004} 
with a typical time step ten times smaller than 
the period of the perturbation $\tau=0.1(2\pi/\omega)$.  
In this work we put a great deal of emphasis in achieving a high degree of accuracy, 
and for this purpose we limit ourselves to relatively small systems.
We consider chains up to $L=24$ sites long, keeping up to $m=1000$ 
states, enough to keep the total truncation error 
(given by the accumulation of the DMRG truncation error and Trotter decomposition) 
below $10^{-8}$.
We have also verified that the behaviour of the pair correlation function 
does not depend qualitatively on the system size. 
Therefore we can assume that our observations will be correct also in 
thermodynamic limit. 

\begin{figure}[h]
\centering \includegraphics[width=8cm]{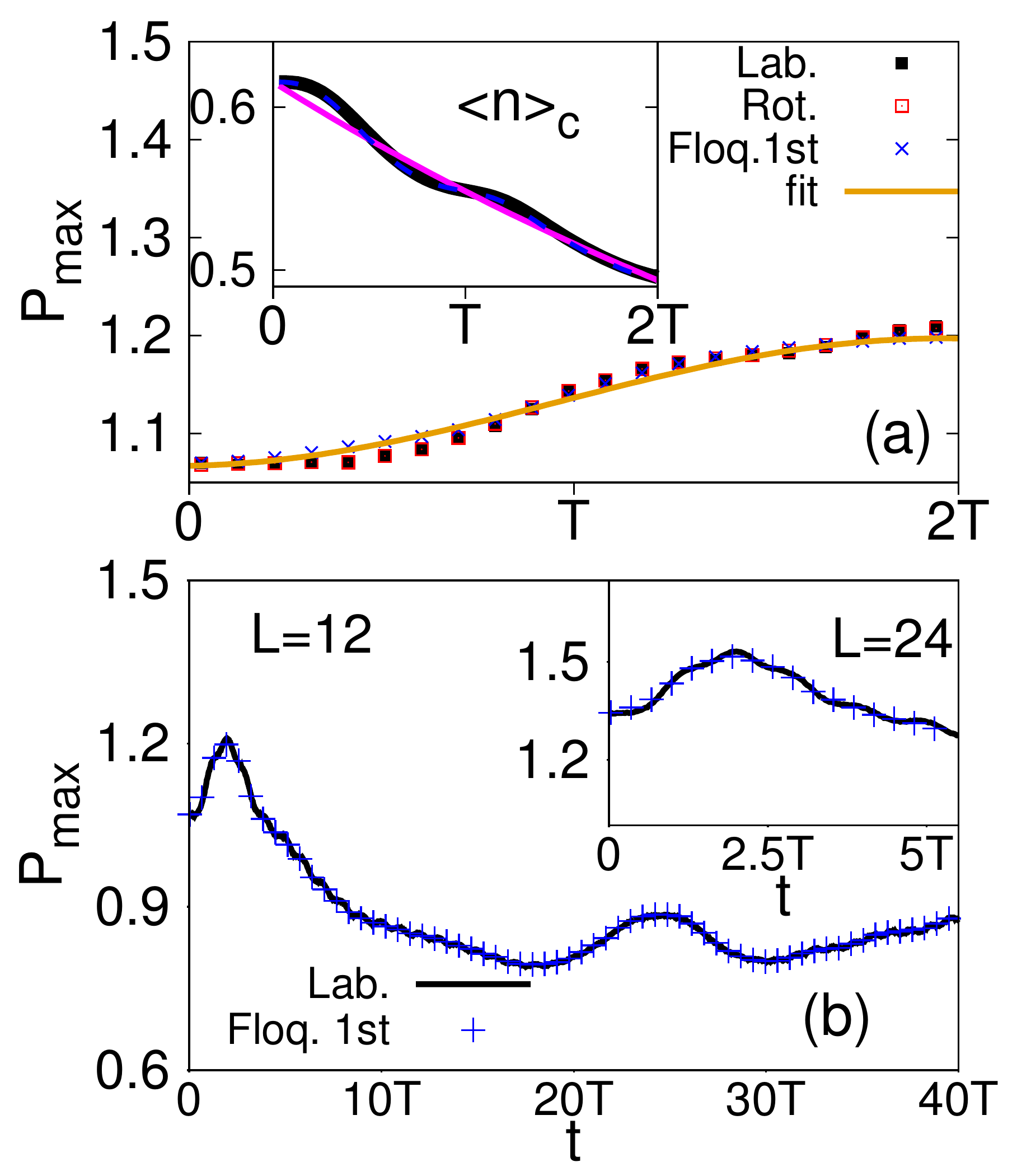}
\caption{(Color online) Panel (a): magnitude of the condensate shown in 
the laboratory, rotating and Floquet first order approximation as a function 
of time. The solid (dark yellow) line indicates the fit
$P_{max}=A+B\cos(Ct+D)$, as discussed in the main text. 
Inset: local density at the center of chain of length $L=12$ as a function 
of time (same filling and other parameters' value as in fig.~\ref{fig1}). Solid thick (black) line indicates data in the laboratory frame, 
the solid thin (magenta) line the data in the zero-order Floquet approximation, 
and the dashed (blue) line the data in the Floquet first order approximation.
Panel (b): Long time behavior of the 
magnitude of the condensate. Notice the agreement between the laboratory frame and the Floquet approximation at first order.
Inset: Magnitude of the condensate as a function of time for a system of length $L=24$, 
keeping up to $m=1000$ DMRG states with a maximum truncation error of $10^{-8}$.} \label{fig1a}
\end{figure}

Fig.~\ref{fig1} shows how, 
for sufficiently small time $t\lesssim 3T$, the momentum of the Fermionic 
condensate, originally at $k=0$, oscillates harmonically around a \emph{finite} 
value $k^{*}$. 
One can fit the motion of the central peak of the condensate in the Brilloin zone
with the relation $k^{*}(t)=2\frac{A}{\omega}(1-\cos(\omega t))$. 
%When $|k^{*}(t)|>\pi$, the value expressing the center of mass momentum of the 
%condensate is reflected of $2\pi$ by periodicity.
Panel (a) of fig.~\ref{fig1a} describes the magnitude of the condensate, $P_{\rm max}(t)=\max_k P(k,t)$, as a function of time
in the four schemes presented in the previous section. 
As it should be the case, the laboratory and rotating frame results
coincide exactly. We also found that the behavior of the condensate peak can be fitted by a relation 
$P_{max}=A+B\cos(Ct+D)$ in the Floquet first order approximation, 
with $C\simeq 1.23|U|$.  
The inset of panel (a) shows the local density at the center of chain of length $L=12$ as a function of time. 
The data in the laboratory frame 
(solid thick black line), which can be considered as an \emph{exact} result, 
is compared with the zero and first order Floquet approximation 
(solid thin magenta and dashed blue lines, respectively). 
%In the limit of small driving amplitude $A/\omega \ll 1$, 
%since $J_{m}(x)\propto x^{m}$ (with $m\neq 0$), one can evaluate an approximate expression 
%for the kick operator, where one retains only the $m=1$ and $m=-1$ terms
%\begin{equation}
%K_{F}^{(1)}(t)\simeq 2i J e^{-i A/\omega} \sin(\omega t)\frac{J_{1}(A/\omega)}
%{\omega}\sum\limits_{\langle i,j\rangle, 
%\sigma}c^\dagger_{i,\sigma}
%c_{j,\sigma} +h.c.. 
%\end{equation}
Even in the large frequency regime $J/\omega =0.05 \ll 1$, 
one can see that the introduction of the kick operator 
at first order in the Floquet expansion is crucial to reproduce 
the out of equilibrium dynamics of the system 
within a single period of the external drive.
%the effect of the kick operator is expected to be small. Indeed, as one can observe 
%in panel (a) fig.~\ref{fig1a}, the introduction of the kick operator 
%at first order in the Floquet 
%expansion gives just a small contribution in the limit of large frequency, $\omega=20$.
Moreover, the results obtained in the laboratory frame 
and the first order Floquet approximation agree
for all the times and system sizes investigated in this work. 
Indeed, panel (b) of fig.~\ref{fig1a} studies the long time behaviour of the magnitude 
of the fermionic condensate for a larger system size chain showing a robust 
agreement between the two schemes. 
In both panels of fig.~\ref{fig1a}, we have retained up to $|m|=5$ harmonics in the expression of the kick operator in eq.~(\ref{eq:kick}).

%We have also verified that our 
%results don't change qualitatively with the system size, as shown in the inset of panel (b) 
%of the same figure.   

\begin{figure}[h]
\centering \includegraphics[width=8.5cm]{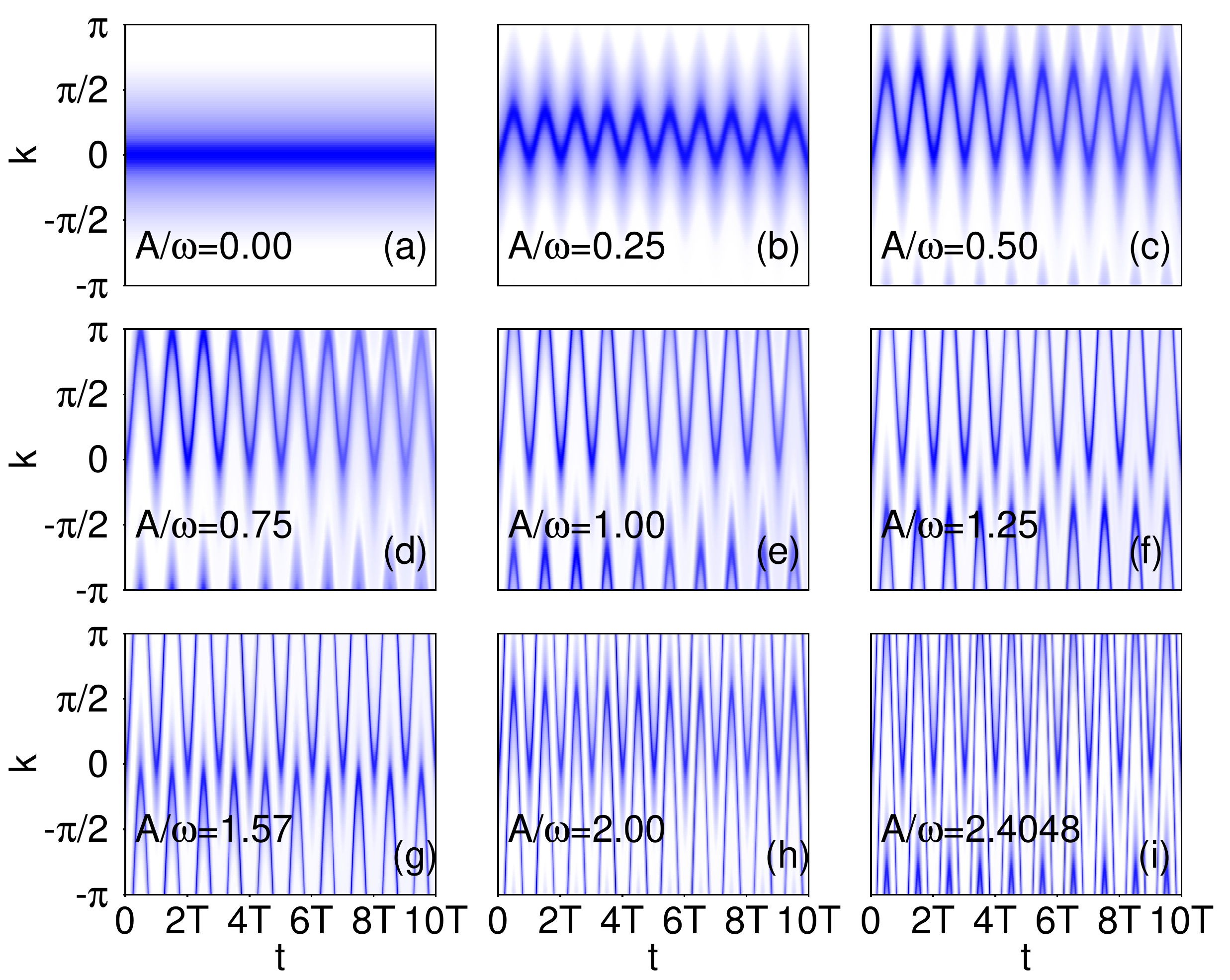}
\caption{(Color online) Pair correlations functions as function of momentum and time
calculated in the laboratory frame for different values of the ratio $A/\omega$.
In this plot, a $L=12$ sites chain has been simulated using $m=400$ 
DMRG states, $U=-4$.} \label{fig2}
\end{figure}

\begin{figure}[h]
\centering \includegraphics[width=8.25cm]{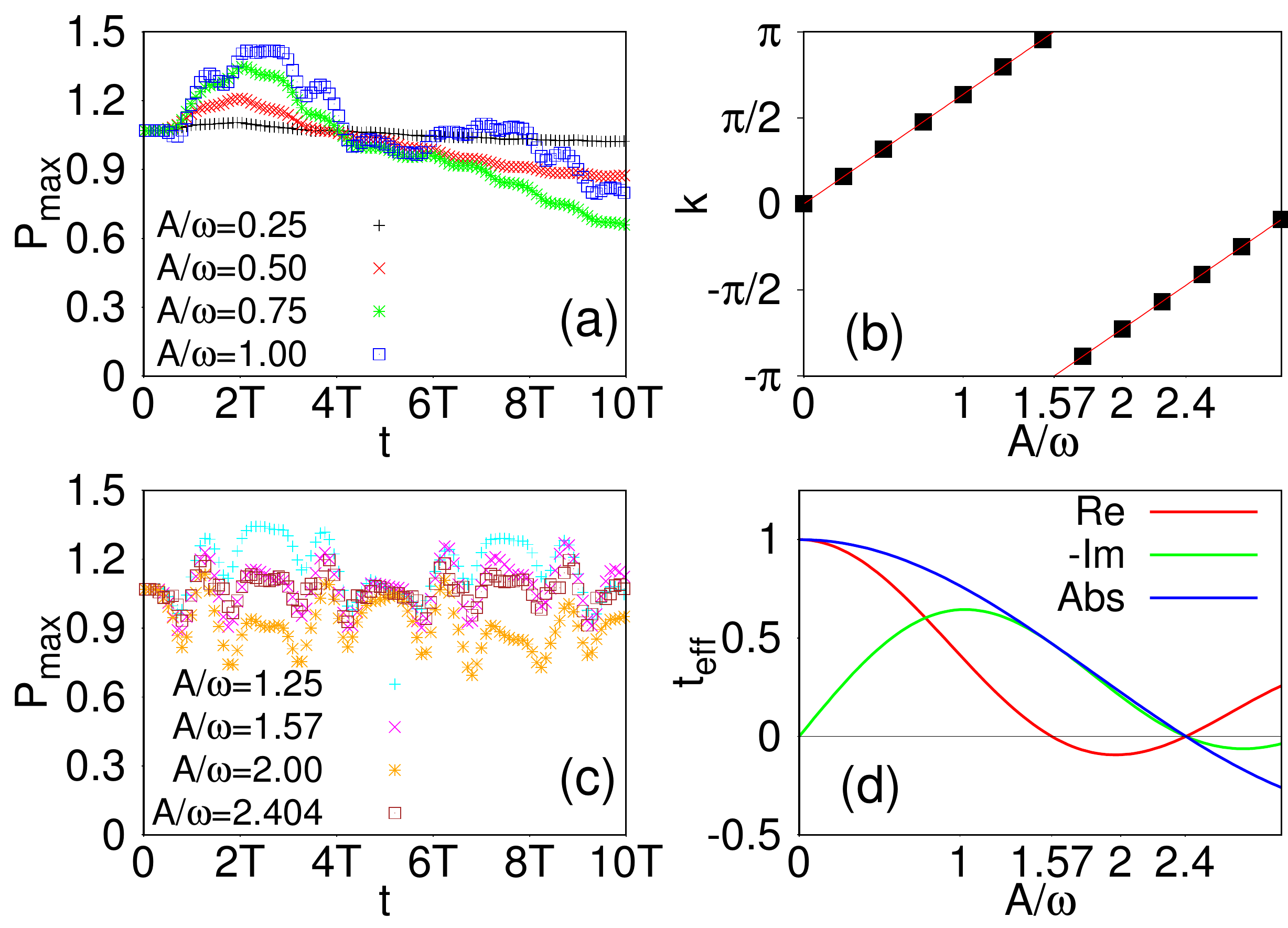}
\caption{(Color online) Panels (a-c): Magnitude of the condensate shown in 
Fig.~\ref{fig2} as a function of time for different values of the ratio $A/\omega$.
Panel (b): 
average position of the center-of-mass momentum of 
the condensate as a function of $A/\omega$ extracted from Fig.~\ref{fig2}. 
The solid (red) line is a guide to the eye. Panel (d):
Solid (red) line represent the real part of the hopping in the Floquet
Hamiltonian in Eq.~(\ref{HF}). Dashed (green) line is the imaginary part, 
short dashed (blue) line indicates its modulus. The line $y=0$ is 
drawn for convenience.} \label{fig2a}
\end{figure}

Fig.~\ref{fig2} shows the pair correlation function for different values of 
the ratio $A/\omega$. One can notice that, by increasing the ratio $A/\omega$, 
the \emph{average} position of the condensate peak moves towards the 
edge of the Brilloin zone. Eventually, it gets reflected at the boundary zone edge
as displayed in Fig.~\ref{fig2a}b, where the average positions of the center
of the condensate for small times $t\lesssim5T$ are shown. By changing the 
driving parameters in the interval $0<A/\omega\lesssim 3$ one can tune the central 
peak of the Fermionic condensate in the entire Brilloin zone (see Fig.~\ref{fig2a}(b)).
This effect demonstrates the possibility of tuning the average momentum 
of the condensate by changing the parameters of 
the external driving. 

Panel (a) of Fig.~\ref{fig2} reproduces the trivial result for no driving as a reference. 
Focusing with more attention on the four panels (b-e), we find (not noticeable by eye) that
the time-averaged position of the condensate in k-space at each period is \emph{not} 
constant in time.
The average momentum of the condensate is slowly reducing its amplitude
as a function of the time. Correspondingly, we have studied the magnitude of the 
center-of-mass of the condensate as a function of time, which
encodes the coherence of the Fermionic condensate. In Fig.~\ref{fig2a}(a) one can notice
that, after an initial transient for times $t\lesssim5T$, 
the coherence of the condensate decreases monotonically. This effect reaches 
its maximum for $A/\omega\simeq0.75$. For larger driving amplitudes, the system builds up
oscillations that survive at longer times, see the blue curve in Fig.~\ref{fig2a}(a).
 Interestingly, for a sufficiently large 
ratio of $A/\omega$, the oscillations of the magnitude of the condensate at large
times are not suppressed anymore but are instead persistent, as shown in Fig.~\ref{fig2a}(c). In order to 
understand this behaviour, it is important to consider the effective description
provided by the Floquet Hamiltonian. 
Fig.~\ref{fig2a}(d) shows real, imaginary and modulus of 
effective hopping parameter, $J_{eff}=J J_{0}(z)e^{-i A/\omega}$, in the Floquet Hamiltonian  
in Eq.~(\ref{HF}). We notice that the modulus of 
the effective hopping represents the well known renormalization of 
the electronic band structure due to the periodic driving, which can result 
in the dynamical fermionic localization at the zeros of the Bessel function $J_{0}(z)$. 
In the interval $1\lesssim A/\omega\lesssim2.5$ the  modulus of the 
effective hopping is significantly smaller than the other  energy 
scale appearing in the Floquet Hamiltonian, 
the Coulomb repulsion $U$, $|J_{eff}| \ll U$. 
Interestingly, in this regime, the suppression of oscillations in the 
magnitude of the  condensate are reduced, as seen in Fig.~\ref{fig2a}(c). 

\begin{figure}[h]
\centering
\hspace*{-0.5cm}
\includegraphics[width=8cm]{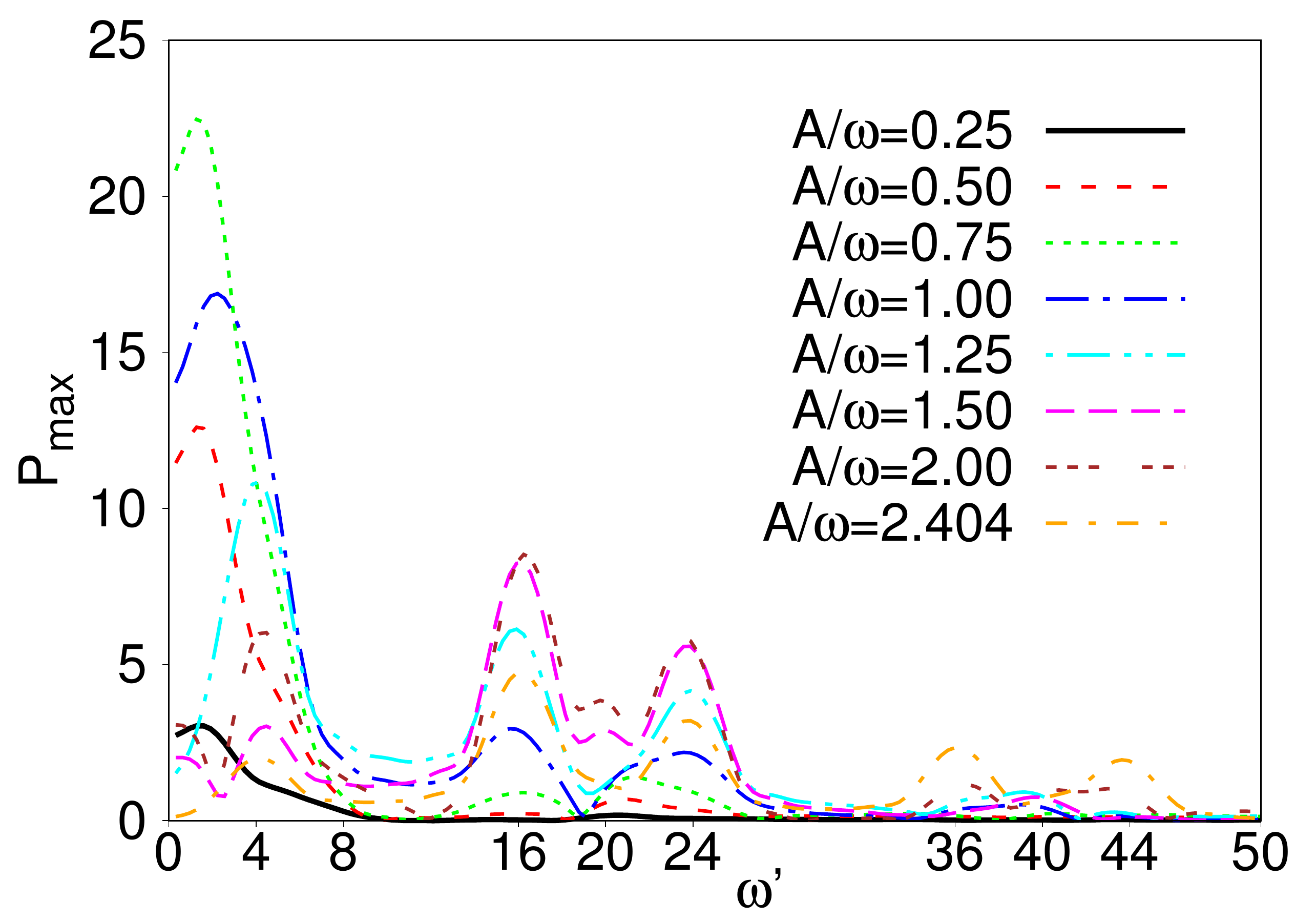}
\caption{(Color online) 
Fourier transform of the curves shown in panel (a-c) of Fig.~\ref{fig2a}.} \label{fig2b}
\end{figure}

Fig.~\ref{fig2b} presents the Fourier transform of the data 
shown in Fig.~\ref{fig2a}(a) and (c). Notice that the curves
obtained for $A/\omega\leq 1$ in Fig.~\ref{fig2a} are not time-periodic
so the corresponding Fourier transforms shown in Fig.~\ref{fig2b} present 
a low frequency peak, which simply reflects this slow decay. For larger driving amplitude $A/\omega=1.0$ one can notice emergence of 
two additional peaks at $\omega\pm U$ in the Fourier spectrum. 
The remaining curves of Fig.~\ref{fig2b} are the Fourier transforms of the 
data shown in Fig.~\ref{fig2a}(c). Same as before, the main feature of the curves 
is the presence of the two peaks at $\omega\pm U$. Notice the appearance of the 
spectral peak at $\omega'\simeq|U|$ together with a small contribution of two peaks at
larger frequency corresponding to higher harmonics, $2\omega\pm U$. The contribution 
of high frequency harmonics is particularly significant at $A/\omega=2.4048$, 
where the absolute value of the effective hopping is zero, see Fig.~\ref{fig2a}(d).
\begin{figure}[h]
\centering \includegraphics[width=8cm]{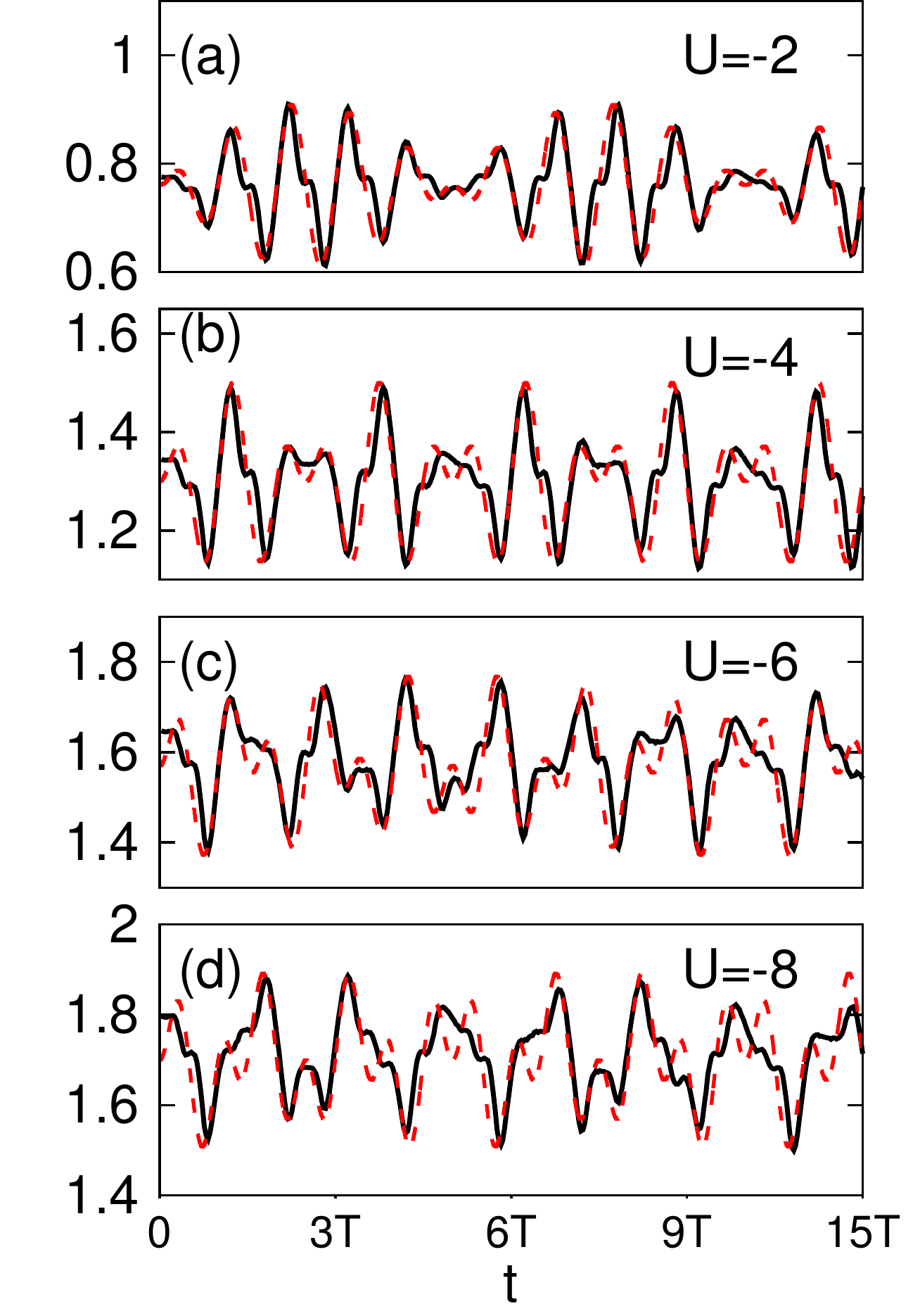}
\caption{(Color online) 
Magnitude of the condensate for $A/\omega=2.4048$ and different values 
of the Coulomb repulsion $U$, as indicated. The red curves represent fits of the 
data according to the relation $P_{max}=A+B\sin(U t)\sin(\omega t)$.} \label{fig3}
\end{figure}
\begin{figure}[h]
\centering \includegraphics[width=8cm]{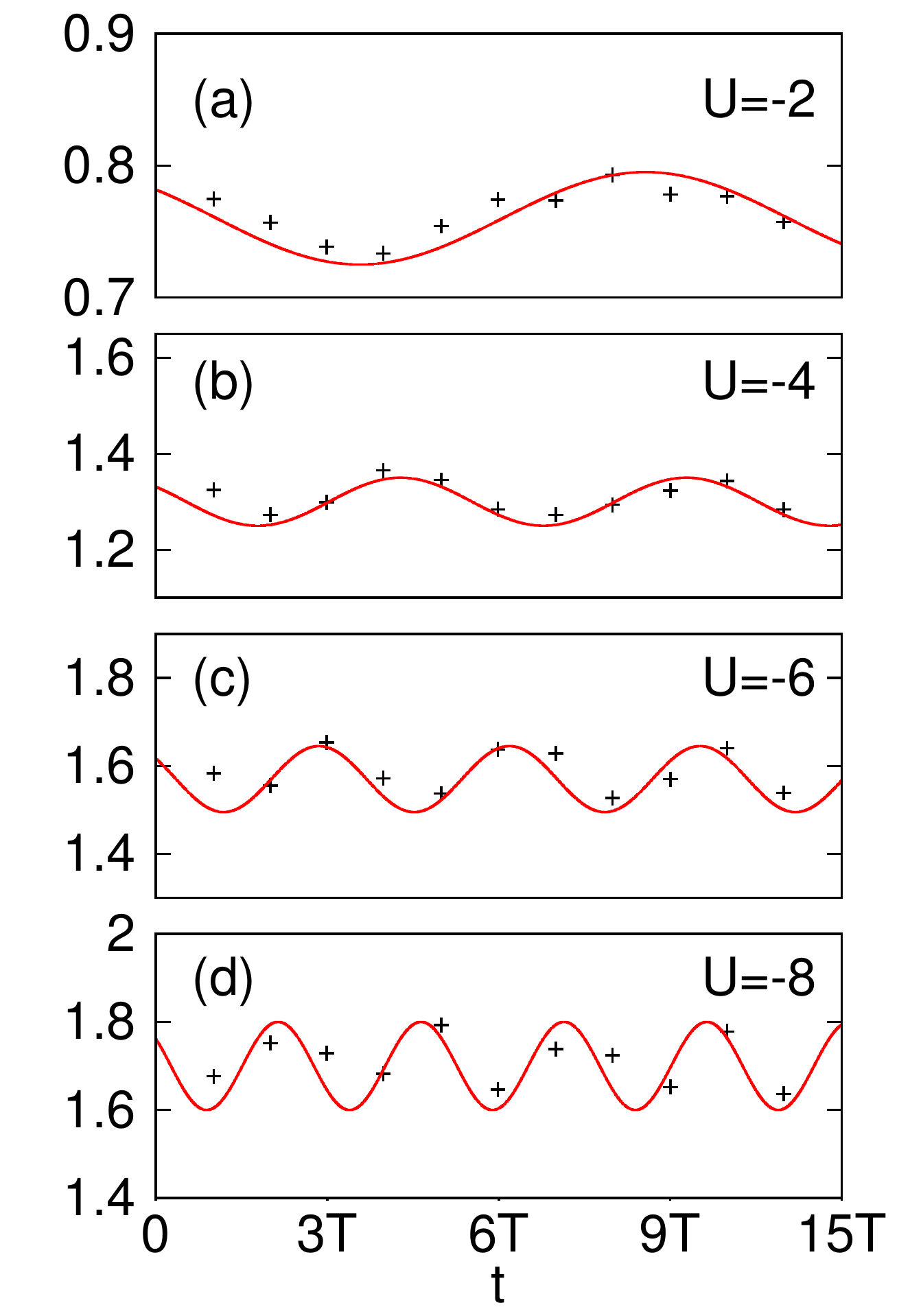}
\caption{(Color online) 
Stroboscopic evolution of the magnitude of the condensate shown in Fig.~\ref{fig3}.
The red curves represent fits of the 
data according to the relation $P_{max}=A-B\sin(U t+\phi)$.} \label{fig3b}
\end{figure}
\begin{figure}[h]
\centering \includegraphics[width=8cm]{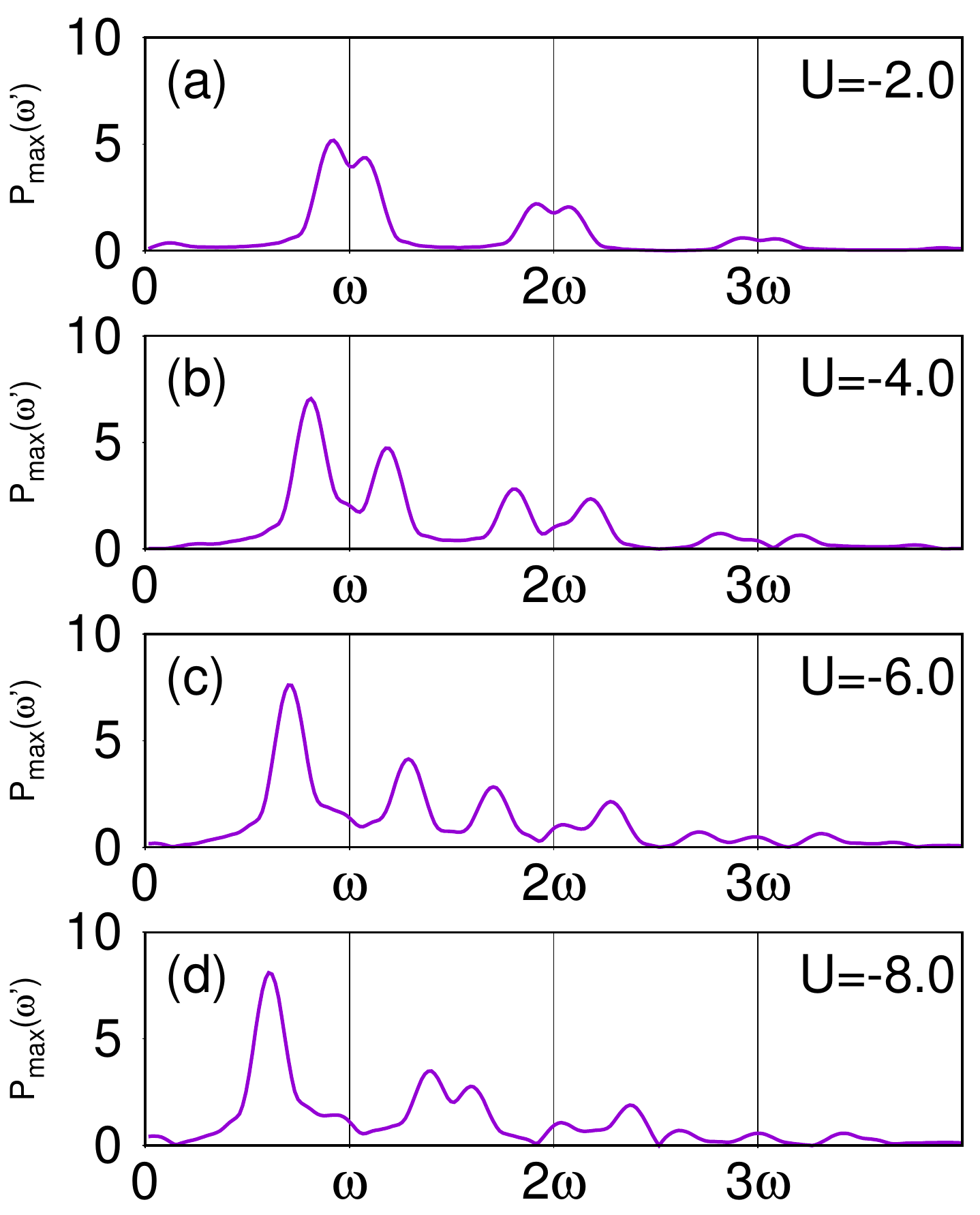}
\caption{(Color online) 
Fourier transform of the curves shown in Fig.~\ref{fig3}.} \label{fig3a}
\end{figure}
The results obtained can be interpreted as in Ref.~[\onlinecite{re:Bukov2016}]. In fact, 
in the regime $U,\omega \gg J$, one can see that in a rotating frame defined by the 
canonical transformation $V(t)=e^{-i (f(t) \sum_{j}j {\hat n}_{j}+Ut\sum_{j}
n_{\uparrow,j}n_{\downarrow,j})}$, the electron-electron interation 
$U$ induces non-trivial phase shifts in the hopping amplitude.
Fig.~\ref{fig3} shows the magnitude of the condensate as a function of time 
for $A/\omega=2.4048$ for different values of the 
Coulomb repulsion $U$. Notice that
an accurate fit of the data can be obtained using the relation 
$P_{max}=A+B\sin(U t)\sin(\omega t)$, showing \emph{beats} with frequencies 
$\omega\pm U$. 
The stroboscopic evolution at times equal to multiples of the period
$T=2\pi/\omega$ of the driving perturbation is shown in Fig.~\ref{fig3b}.
One can clearly observe that in a non-stroboscopic analysis, both the fast
(the frequency of the driving force $\omega=20$) and the slow (the Coulomb repulsion $U$) 
time scales of the dynamics are resolved. 
A Fourier analysis of the data provides more details on the non-stroboscopic evolution, 
see Fig.~\ref{fig3a}. We find a clear contribution of at least
three harmonics, with pairs of peaks at $n\omega\pm U$ (with $n=1,2,3$).
For each pair, the distance between the two peaks is proportional to $2U$. 
The intensity of the Fourier peaks decreases almost exponentially
with the order of the harmonics.
Indeed, the zero-order Floquet expansion
presented in the previous section for $A/\omega=2.4048$
corresponds to a Floquet Hamiltonian given just by the interaction term
$H_{F}^{(0)}\simeq U \sum\limits_{i} n_{i,\sigma}n_{i,\bar{\sigma}}$. 
Therefore, this confirms that 
the zero-order Floquet approximation governs the stroboscopic long 
time evolution of the system.

In the next section, we will study the properties of the pairing states of the 
system in the laboratory reference frame by using different 
quench protocols.

\section{Characterization of the pairing phase: quenching protocols}

\begin{figure}
\centering \includegraphics[width=8.6cm]{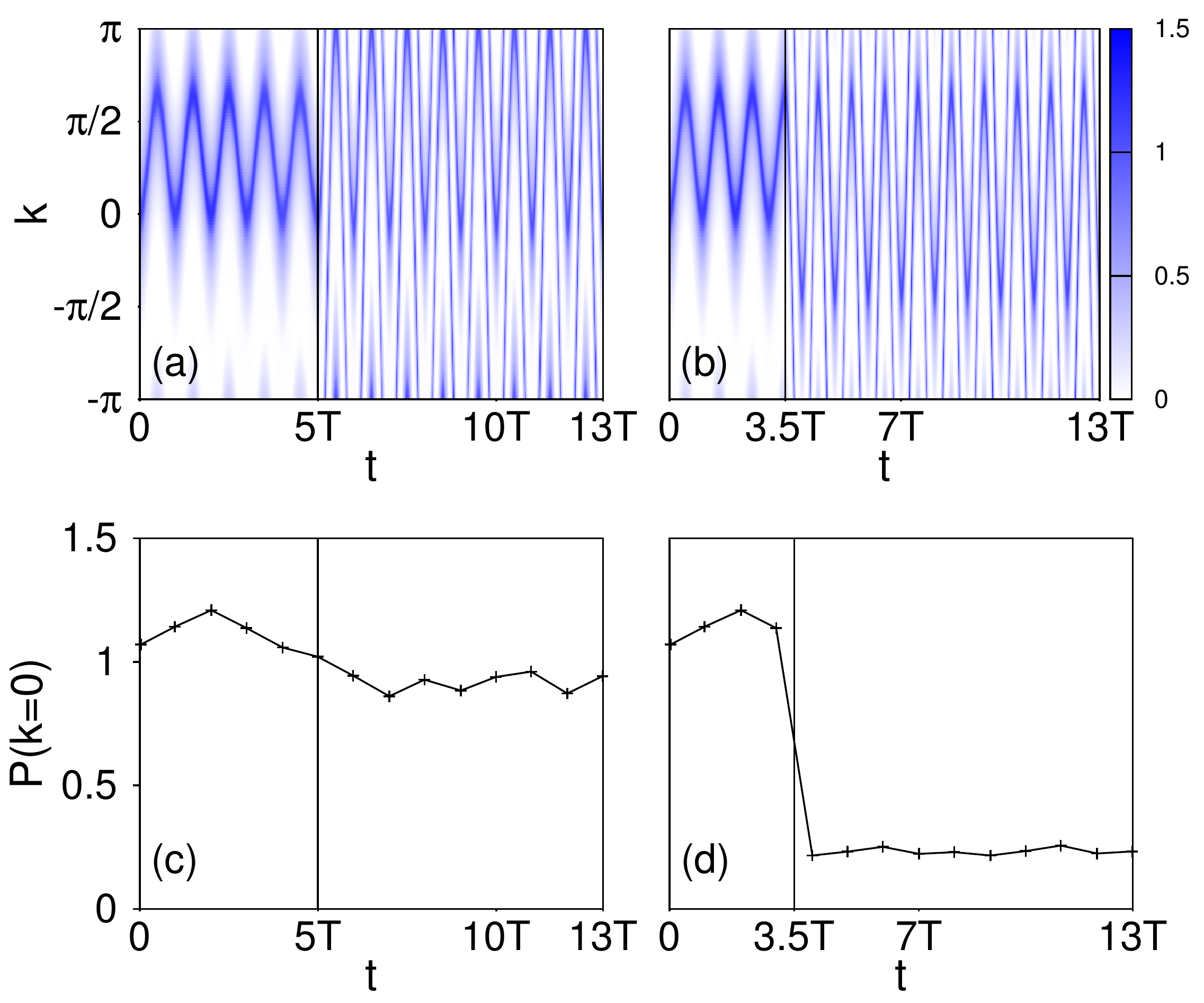}
\caption{(Color online) Panels (a)-(b): Pair correlation functions before and after 
a quench of the amplitude of the driving from $A_1/\omega=0.5$ to $A_2/\omega=2.4048$.
In panel (a) the quench is performed at $t=5T$, while in panel (b) at $t=3.5T$. 
Panels (c)-(d): Intensity of the peak at zero momentum of the same functions shown 
in panels (a) and (b) evaluated at each period of time $T$. In this plot, a $L=12$ sites chain 
has been simulated using $m=400$ DMRG states.} \label{fig4}
\end{figure}

In this section we investigate the properties of the pairing phase by abruptly changing  
either the interactions, or the parameters of the periodic driving force. 
We first consider the quench protocol where we abruptly
 change the ratio $A/\omega$ of the driving
 but leaving the bare hopping and the $U$
attraction strength fixed. In practice, we change the values of the amplitude $A$ and keep
the frequency of the driving fixed. In Fig.~\ref{fig4}, we assume that before the quench
$A_1/\omega=0.5$ while after the
quench $A_2/\omega=2.4048$, which is the ratio corresponding to coherent destruction of 
tunnelling. In panels (a) and (b), the pair correlations functions are shown as a
function of the momentum $k$ and the time $t$. In the panels (c) and (d) the magnitudes 
of the peaks at zero momentum are shown for $t=nT$, with $n$ non-negative integer. 
In panels (a) and (c), as also indicated by a vertical solid line, the quench is
performed at $t=5T$, that is after exactly $5$ periods of the driving force. 
In panel (b) and (d), the quench is performed after $3.5$ periods of the driving.
Panels (a) and (c) show a very interesting behaviour. If, in a real experiment, one is
able to quench the amplitude of the driving at an instant of time that is an exact multiple of the driving
frequency $\omega$, one would measure, in average, just a shift in the 
central peak of the pair distribution function, with no destruction of the condensate. 
On the contrary, if a quench of the amplitude of the driving is done in any instant of
time between $nT$ and $(n+1)T$ (we picked $3.5T$ as quenching time just for
convenience), the oscillation of the peak of the condensate picks up a \emph{random}
phase factor which determines a significant (about 80\%) suppression of the hight of the 
condensate peak (see panel (d)). 

\begin{figure}
\centering \includegraphics[width=8.5cm]{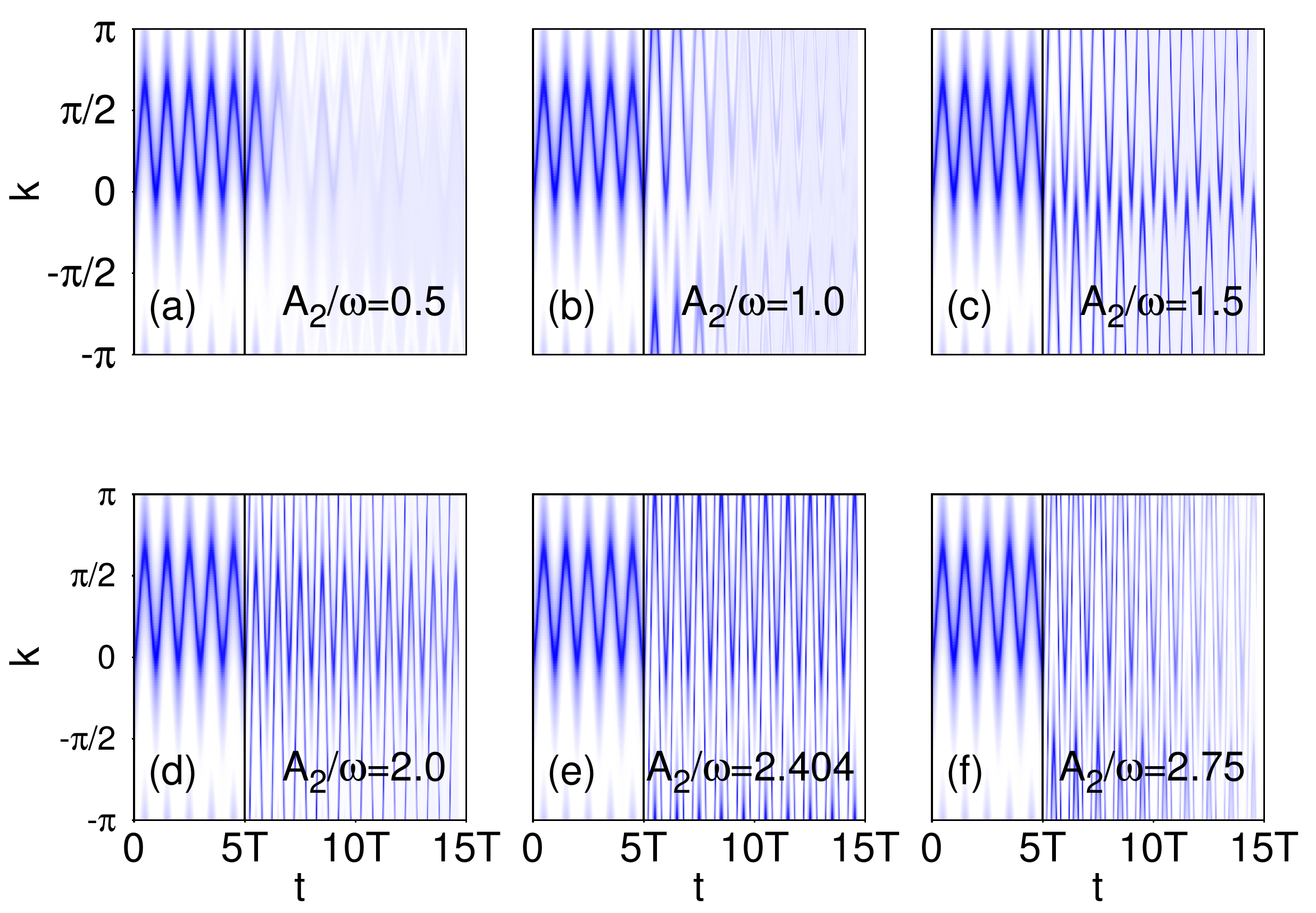}
\caption{(Color online) Pair correlations functions as function of momentum and time
for different values of the ratio $A_2/\omega$. A quench in the interaction from $U_{1}=-8$ to $U_{2}=0$ is performed at $t=5T$. For $t<5T$, $A_1/\omega=0.5$.
In this plot, a $L=12$ sites chain has been simulated using $m=400$ DMRG states.} \label{fig5}
\end{figure}

We now consider the possibility of quenching the interaction between the fermionic 
atoms loaded on the optical lattice. In panel (a) of Fig.~\ref{fig5}, 
we assume that the fermionic condensate is initially prepared with $U_1=-8.0$ and with driving
 parameters given by the ratio $A_1/\omega=0.5$, such that the center of the condensate
  is around $k^{*}=1$ in average.
At $t=5T$, we quench the interaction from $U=-8$ to zero, keeping 
the parameters of the
driving fixed. After about a period of oscillation, the fermionic condensate is
almost completely suppressed, because of the lost attraction between the fermions. 

\begin{figure}
\centering \includegraphics[width=8cm]{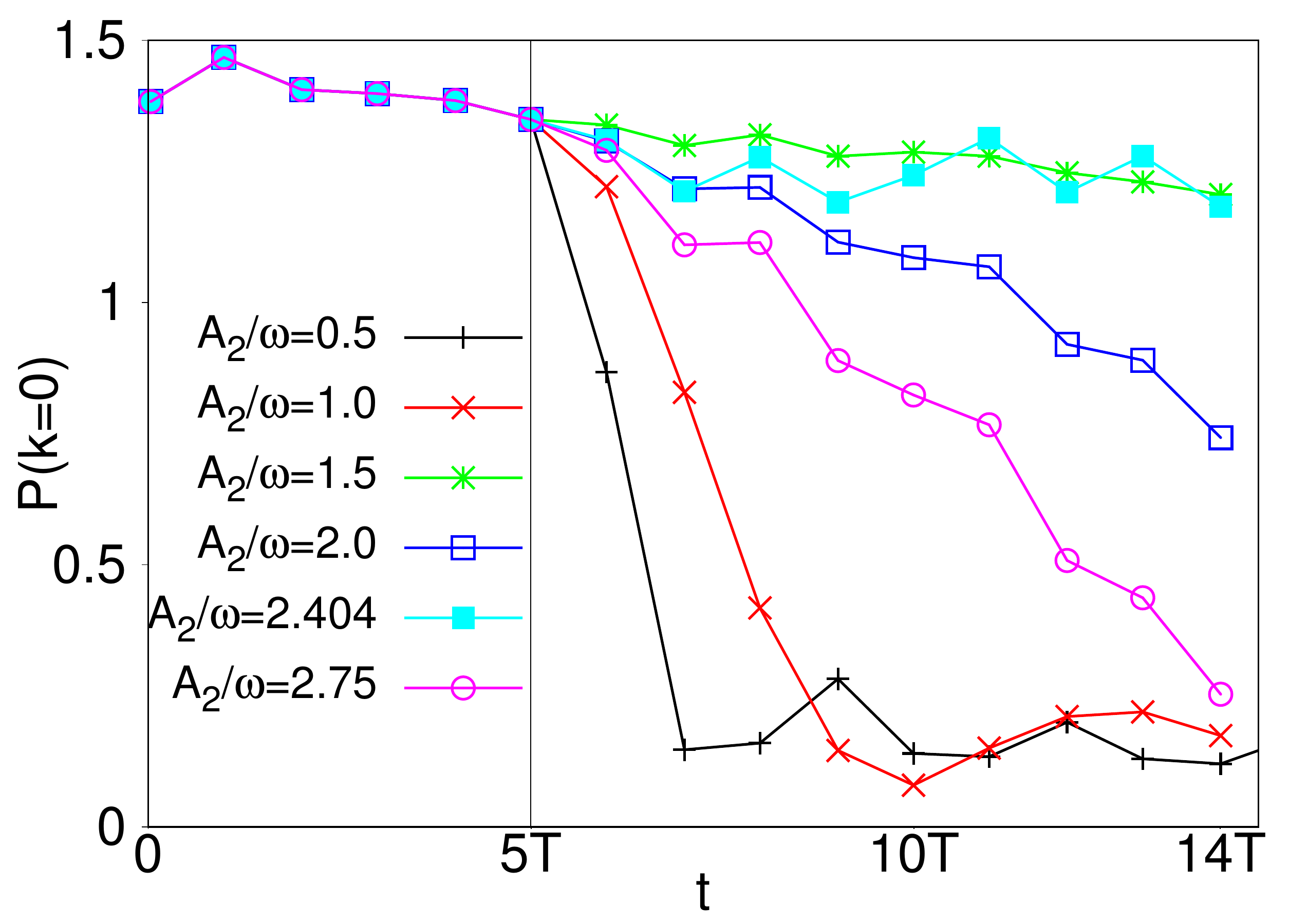}
\caption{(Color online) Zero momentum contribution of the Pair correlations
 functions shown in Fig.~\ref{fig4} evaluated at each period of time $nT$, 
with $n=0,1,2,..,14$. In this plot, $A_1/\omega=0.5$ and $U_{1}=-8.0$.} \label{fig6}
\end{figure}

\begin{figure}
\centering \includegraphics[width=8.5cm]{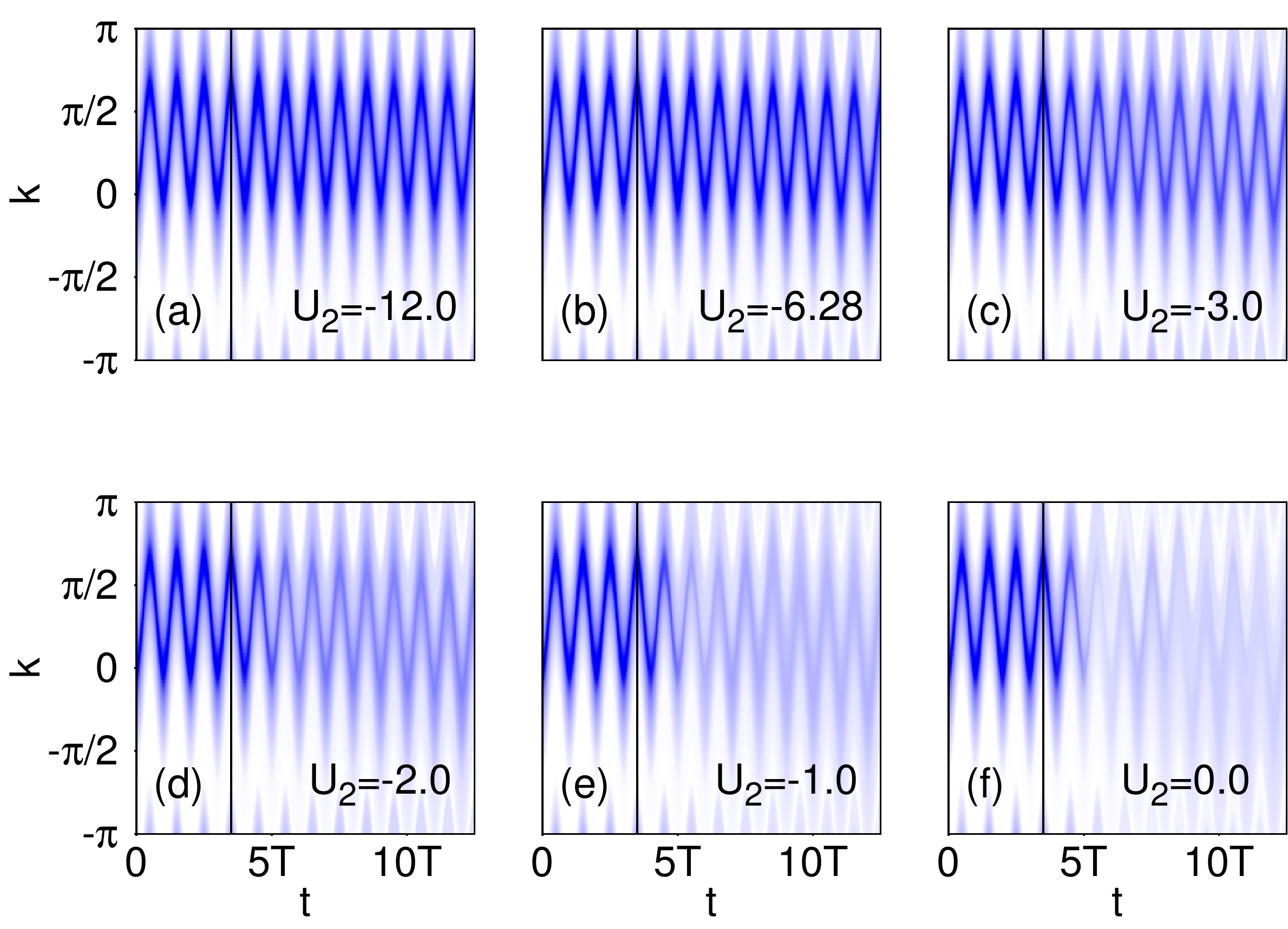}
\caption{(Color online) Pair correlations functions before and after a quench at
$t=3.5T$ of the interaction $U$. Initially, $U_1=-4.0$, 
while driving parameters are given by the ratio $A/\omega=0.5$.
In this plot, a $L=12$ sites chain has been simulated using $m=400$ DMRG states.} \label{fig7}
\end{figure}

A very non-trivial effect is visible if one increases the ratio $A_2/\omega$ after 
the quench: the coherence of the condensate displays revivals 
even if the bare interaction between the fermions is quenched to zero. 
A stroboscopic measure of the coherence of the condensate is provided in the 
panel (a) of Fig.~\ref{fig6}. The revival of the coherence of the condensate is 
maximum around $A_2/\omega\simeq\pi/2$ and $A_2/\omega\simeq2.4048$. All the other cases
plotted, $A_2/\omega=1$, $A_2/\omega=2$, and $A_2/\omega=2.75$ show a decrease of 
the magnitude of the condensate peak. 
A qualitative explanation of the effects goes as follows: at the
time of the quench, the effective Floquet Hamiltonian of the system changes
accordingly with the new parameters of the driving. As demonstrated in Fig.~\ref{fig2a}(d), 
the real part of the effective hopping vanishes exactly at $A/\omega\simeq\pi/2$ 
and $A/\omega\simeq2.4048$. As shown in Ref.~[\onlinecite{re:Creffield2011}] the real part of the effective
hopping is proportional to the spreading of a condensate prepared in real space,
which is then let evolved in time under the action of the driving. 

We finally address a peculiar quenching protocol where we keep the
parameters of the driving fixed and, after preparing the condensate with a finite
average center-of-mass momentum, we quench the attraction strength 
at time $t=3.5T$ (as indicated by the black vertical line).
Results are shown in Fig.~\ref{fig7}.
If the strength of the attraction between fermions is quenched to a value which is 
larger in modulus than the initial value $U=-4$ (see panels for $U_2=-12$ and $U_2=-6.28$), the
center-of-mass momentum does not change, while the magnitude of the condensate peak
is increased (not shown in the plot). If the modulus of the strength of the attraction between the
fermions is decreased, not only the 
coherence of the condense decreases, going to zero if $U_2=0$, but the average 
center-of-mass momentum also drifts toward zero (see panels for $U_2=-2$ 
and $U_2=-1$). 
  
\section{CONCLUSIONS}
In this work, we have studied the one-dimensional attractive Fermionic 
Hubbard model subjected to a sinusoidal periodic drive. 
We have assumed that the frequency of the driving force is large enough such 
that the system is in the off-resonant regime, and Floquet theory can be applied.
To benchmark our numerical simulations, we have verified that a Van Vleck high-frequency expansion 
in the rotating frame truncated to first order is able 
to describe very well the numerical results obtained with time dependent 
DMRG. In particular, at zero order, 
the stroboscopic dynamics in a rotated reference frame is described by 
an effective Floquet Hamiltonian with a complex phase in the hopping term 
which introduces a shift of the condensate center-of-mass momentum. 
We have also derived the \emph{micromotion} or \emph{kick} 
operators and verified that their inclusion at first order of the 
high-frequency expansion is crucial to understand 
the dynamics of the system within one period of the external drive.

By tuning the amplitude and the 
frequency of the driving force, one can realize an unconventional 
Fermionic condensate with a finite center-of-mass momentum. 
This pairing state is similar to a FF state, 
because it represents a coherent matter field of Cooper-like pairs 
at finite momentum.
In particular, the center-of-mass of the condensate can be tuned at 
will across the Brilloin zone.
Moreover, we have shown that the coherence of the condensate, which is 
encoded in the magnitude of the pair correlation function in momentum 
space, can be enhanced by tuning the amplitude and 
frequency of the drive to ratios close to those where one 
obtains fermionic dynamical localization. 

From the analysis of the real time evolution of the magnitude of the 
condensate, we found that the stroboscopic dynamics is governed by 
oscillations with period proportional to the electron-electron 
attraction strength $U$, while the full time evolution presents 
also oscillations given by the driving frequency $\omega$, creating 
\emph{beats} at $\omega\pm U$. 
 
Finally, we show that using different 
quenching protocols, the condensate can be \emph{frozen} 
at a particular finite center-of-mass momentum and coherence.
Moreover, the coherence of the condensate displays revivals 
even if the bare interaction between the fermions is quenched to zero.
These protocols are within experimental reach.

Our work could pave the way for engineering out-of equilibrium FFLO-like phases 
in multicomponent Fermi gases without population imbalance.

\section{ACKNOWLEDGMENTS}
This work was conducted at the Center for Nanophase
Materials Sciences, sponsored by the Scientific User Facilities 
Division (SUFD), Basic Energy Sciences (BES), U.S. Department of Energy (DOE), 
under contract with UT-Battelle. A.N. acknowledges support by the Center for 
Nanophase Materials Sciences, and by the
Early Career Research program, SUFD, BES, DOE.
A.E.F. acknowledges the DOE, Office of Basic Energy
Sciences, for support under grant DE-SC0014407. A.P. was supported by
NSF DMR-1506340, ARO W911NF1410540 and AFOSR FA9550-16-1-0334.

\bibliography{biblio}
\end{document}